\documentclass[onecolumn,pra, superscriptaddress,longbibliography,margin=1mm 12pt,,colorlinks=true,urlcolor=blue,citecolor=cyan,linkcolor=cyan]{revtex4-2} 
\usepackage{bm}
\usepackage{hyperref}
\usepackage{graphicx}
\usepackage{amsmath}
\usepackage{csquotes}
\makeatletter
\newcommand{\Autoref}[1]{\begingroup
  \def\figureautorefname{Figure}%
  \def\tableautorefname{Table}%
  \def\sectionautorefname{Section}%
  \autoref{#1}%
\endgroup}
\makeatother
\makeatletter
\def\figureautorefname{figure}     % "figure" 小文字
\def\tableautorefname{table}       % "table" 小文字
\makeatother
\begin{document}

%%%% Article title to be placed here
\title{Drawing with water waves}
\date{\today}
%\author{%%%% Author details
%Taiga Kanehira$^{1}$, James Steer$^{2}$
%Laura-Beth Jordan$^{3}$, Shaun Fraser$^{3}$, Thomas Davey$^{3}$,
%Hidemi Mutsuda$^{1}$ and Samuel Draycott$^{5}$}

%%%%%%%%% Insert author address here
%\address{$^{1}$Graduate School of Advanced Science and Engineering, Hiroshima University, Hiroshima, Japan\\
%$^{2}$European Centre for Medium Range Weather Forecasts, Reading, UK\\
%$^{3}$School of Engineering, FloWave Ocean Energy Research Facility, Institute for Energy Systems,University of Edinburgh\\
%$^{5}$School of Engineering, University of Manchester, Manchester, UK}
\author{Taiga Kanehira}
\email{taiga.kanehira@gmail.com}
\address{Graduate School of Advanced Science and Engineering, Hiroshima University, Hiroshima, Japan}
\author{James Steer}
\address{European Centre for Medium Range Weather Forecasts, Reading, UK}
\author{Laura-Beth Jordan}
\address{School of Engineering, FloWave Ocean Energy Research Facility, Institute for Energy Systems,University of Edinburgh, UK}
\author{Shaun Fraser}
\address{School of Engineering, FloWave Ocean Energy Research Facility, Institute for Energy Systems,University of Edinburgh, UK}
\author{Thomas Davey}
\address{School of Engineering, FloWave Ocean Energy Research Facility, Institute for Energy Systems,University of Edinburgh, UK}
\author{Hidemi Mutsuda}
\address{Graduate School of Advanced Science and Engineering, Hiroshima University, Hiroshima, Japan}
\author{Samuel Draycott}
\address{School of Engineering, University of Manchester, Manchester, UK}

%%%% Subject entries to be placed here %%%%
%\subject{fluid mechanics, ocean engineering, computational mechanics}

%%%% Keyword entries to be placed here %%%%
%\keywords{wave field manipulation, wave generation, surface gravity waves, computational fluid dynamics, smoothed particle hydrodynamics}

%%%% Insert corresponding author and its email address}
%%\corres{Insert corresponding author name\\
%\email{taigakanehira@hiroshima-u.ac.jp}}

%%%% Abstract text to be placed here %%%%%%%%%%%%
\begin{abstract}
The deterministic reproduction of complex 3D wave fields remains a significant challenge in ocean engineering. This study proposes a novel methodology for \enquote*{drawing} arbitrary 2D curves and 3D volumetric shapes on a water surface using transient multi-directional focused waves. To overcome the limitations of conventional discrete-point focusing methods, our framework integrates Bézier curve parametrisation, equal arc-length sampling, and an Iterative Amplitude Correction (IAC) algorithm. This effectively mitigates wave height overshoot and enables precise spatial superposition of spectral components.

The method's effectiveness was validated through linear wave theory and Smoothed Particle Hydrodynamics (SPH) simulations, successfully reproducing 2D characters and a 3D human face. Physical experiments in the FloWave circular wave basin further demonstrated target shape generation, such as a 2D star and 3D pyramid. Although further integration of nonlinear wave theories is necessary for high-amplitude accuracy, this technique establishes a deterministic methodology for creating arbitrary 2D and 3D surface geometries. It represents a significant advantage in wave field control for ocean engineering applications.
%\absbreak 

\end{abstract}
%%%%%%%%%%%%%%%%%%%%%%%%%%%
\maketitle
%\rsbreak

%%%%%%%%%% Insert the texts which can accomdate on firstpage in the tag "fmtext" %%%%%

% --------------------------------------------
\section{Introduction}
% --------------------------------------------
The precise manipulation of the water surface is of critical importance in ocean engineering, where the reproduction of complex three-dimensional wave fields is essential for evaluating the structural integrity and survivability of offshore structures and systems. Although this technological need has parallels in biotechnology e.g., where Faraday waves are used as templates to organise biological cells via steady-state resonance at micro to millimetre scales governed by surface tension \cite{naseer2017, Petta_2021, Rasouli2023}, ocean engineering primarily addresses the more complex challenge of transient wave phenomena at the macro-scale, where gravity and inertia dominate. Unlike static, periodic patterns formed through continuous excitation in bio-applications, the reproduction of arbitrary spatial profiles of gravity waves requires precise transient focusing of wave energy to create non-periodic, focused waves at a specific instant \cite{naito2005, ohmatsu2009}. This transition from static resonance to deterministic transient shaping demands a significantly higher level of phase synchronisation and amplitude control. Consequently, establishing a methodology that can \enquote*{draw} arbitrary 2D/3D shapes as transient events remains a challenging task, bridging the gap between fundamental fluid mechanics and advanced offshore engineering applications.

The necessity for such deterministic 3D control is underscored by the inherent complexity and hazards of natural sea states. In the open sea, offshore ocean waves are characterised by the superposition of multi-directional wave systems, forming highly complex 3D wave fields. Previous reports have identified that the interaction between two wave systems, such as wind waves and swells, often leads to the occurrence of abnormal waves \cite{Cavaleri2012,Trulsen2015,Zhang2017,McAllister_2019}: a phenomenon that has sometimes resulted in marine accidents and loss of life \cite{Tamura2009,Cavaleri2012,Trulsen2015,Zhang2017}. Recent studies have highlighted that as energy spreading and crossing angles in these multi-directional fields increase the waves tend to exhibit standing wave behaviour \cite{McAllister2024}. It has been demonstrated that such focused waves can reach a maximum steepness up to four times greater than that of equivalent two-dimensional (2D) waves  (see e.g. \cite{Stokes1880}) at their breaking onset \cite{McAllister2024}. Therefore, to establish a reliable design methodology for offshore structures, it is an urgent task to develop a technique that can intentionally and precisely reproduce 3D wave profiles representative of extreme events (i.e. not constrained by spatial focusing of parametric spectra) within a controlled laboratory environment.

Despite the clear necessity for reproducing such complex sea states, research on theories to intentionally design and generate arbitrary-shaped 3D focused waves is still limited.
Most existing studies have primarily focused on drawing arbitrary patterns in a horizontal plane as a demonstration of control precision.
Naito et al. conducted a pioneering study on water wave drawing \cite{naito2005}. 
They expanded a planar function $\zeta(r, \theta)$ into a Dini series. 
The spatial irregularities of the water surface were formulated as a sum of Bessel functions. 
This approach treated the surface as a superposition of cylindrical waves. 
Consequently, an arbitrary function is expressed as a series sum of two components:
One is the azimuthal direction, representing the mode number of the Bessel function, and the other is the radial direction, representing the frequency of the component waves. 
They validated this method using the circular AMOEBA tank \cite{Naito1999}, which features 50 wavemakers around the tank's circumference. 
They successfully reproduced characters such as the letter \enquote*{S}. 
However, the finite number of wavemakers restricted the azimuthal series components. 
This limitation makes it difficult to reproduce sharp angles or asymmetric lines, such as the letter \enquote*{K}.

Ohmatsu addressed this issue by adopting the \enquote*{time-reversal method,} in which arbitrary characters are represented as a dot matrix \cite{ohmatsu2009}. This approach involves generating focused waves at specific coordinates to act as individual dots. To achieve this, they mathematically derived the impulse response function for a flap-type wavemaker in an infinite domain subjected to a $\delta$-function velocity. This response is characterised by frequency components that increase over time $t$. By inputting the time-reversed signal---generating high frequencies first, followed by lower ones---wave dispersion causes energy to concentrate at a target point. Consequently, this method can generate focused waves at any location within 2D or circular basins. This allows specific shapes to be constructed as sets of discrete points. The theory was validated in the Deep-sea basin at the National Maritime Research Institute using 128 multi-segmented wavemakers \cite{Maeda2004}, successfully reproducing characters such as \enquote*{NMRI}\cite{ohmatsu2009}.

Nevertheless, these existing methods have limitations in their application to arbitrary shapes. 
First, the dot-based method by Ohmatsu et al. lacks spatial density for smooth, continuous curves. 
Second, a problem arises when multiple focused waves are generated in close proximity. 
Local constructive interference of wave energy causes the wave height to greatly exceed the target. 
This leads to the premature collapse of the wave profile. 
Third, previous research remained limited to 2D \enquote*{lines} or \enquote*{characters} on the water surface. 
There are no recorded cases of physically reproducing 3D volumetric wave shapes.
This study develops a new method to overcome the limitations of these previous works and enables the reconstruction of precise 2D curves and 3D volumetric shapes on the water surface. 
The primary novelty of this paper consists of the following four points:

\begin{enumerate}
    \item \textbf{Focused wave generation through spatial focusing of spectral components:} 
    Instead of the Dini expansion proposed by Naito et al., we achieve focussing by precisely aligning the phases of numerous spectral components from diverse angles. Furthermore, the wavemaker signals are generated using the circular wave tank theory established by \cite{GYONGY2014329}, rather than relying on the time-reversal method in \cite{ohmatsu2009}. This approach allows intuitive manipulation of the wave spectrum, enabling more precise and flexible control over the focused wave profile.

    \item \textbf{Introduction of Bézier curves and equal arc-length sampling:} 
    Target contours are mathematically defined using Bézier curves. 
    We introduce an \enquote*{equal arc-length interval} method to place evaluation points evenly, which prevents waves from collapsing due to evaluation points clustering at high-curvature regions. 
    This enables the drawing of extremely complex and smooth 2D figures.

    \item \textbf{Iterative Amplitude Correction (IAC) algorithm:} 
    To address wave height overshoot caused by the interference of adjacent focused waves, we developed a new iterative algorithm. This approach evaluates the error between the target and superimposed wave heights, feeding back correction factors until the surface converges to the desired shape. This is done rapidly using linear theory before being used to define wavemaker inputs in experimental or numerical tanks.

    \item \textbf{Extension to 3D volumetric shapes:} 
    We established a theory to generate complex 3D surfaces, including pyramidal shapes and the topographical features of a human face. The latter, extracted from 3D CAD models, demonstrates the robustness and versatility of the proposed method.
\end{enumerate}

The remainder of this paper is organised as follows. \S \ref{Sec:Methodology} details the proposed methodology, including the focused wave generation, the Iterative Amplitude Correction (IAC) algorithm, and the experimental setup. \S\ref{Sec:Results} presents the results and validation for both 2D and 3D wave drawing through linear theory, SPH simulations, and physical experiments. Finally.  \S\ref{Sec:Conclusions} provides concluding remarks.

% --------------------------------------------
\section{Methodology}
\label{Sec:Methodology}
% --------------------------------------------

\subsection{Concept of Wave Drawing and Amplitude Correction Approach}
\begin{figure}[!h]
\centering\includegraphics[width=0.5\textwidth]{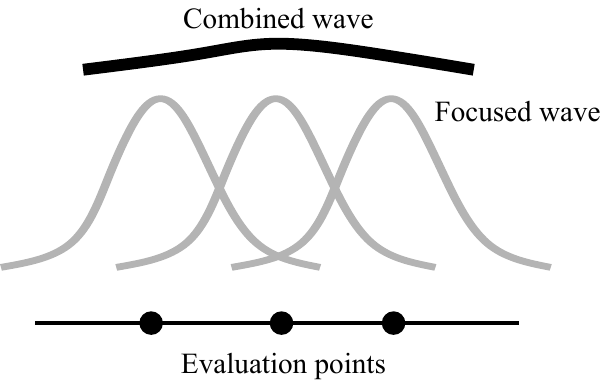}
\caption{Concept of wave shape design based on the superposition of focused waves, where individual focused waves act as component waves to form the final spatial wave pattern.}
\label{fig_concept}
\end{figure}

In this section, we present our proposed methodology for generating target wave shapes. \autoref{fig_concept} illustrates the fundamental concept, where the spatial wave pattern is constructed as a superposition of focused wave crests. In this case, the focused waves are generated using a circular wave tank, where multiple wavemakers are arranged around the tank to produce waves converging towards a focal point. By carefully adjusting the amplitudes and phases of individual focused waves, the overall water surface elevation can be controlled to achieve a desired wave profile. The mechanism for generating focused waves is described in \S\ref{sec:focused_wave}, followed by an explanation of the arrangement of evaluation points for both 2D wave lines and 3D wave shapes in \S\ref{sec:2d_approach} and \S\ref{sec:3d_approach}. Finally, in \S\ref{sec:Amp_correction}, we introduce an amplitude correction approach to adjust the final wave amplitude.

% ----------------------------------------------------
\subsubsection{Generation of Focused Wave}
\label{sec:focused_wave}
% ----------------------------------------------------
A regular wave (a single plane linear wave) can be expressed in exponential form as:
\begin{equation}
\eta(\mathbf{x}, t) = a e^{i \theta} + \text{c.c.}
\end{equation}
where, $\eta(\mathbf{x}, t)$ represents the surface displacement, $\mathbf{x} = (x, y)$ is the position vector, $a$ is the wave amplitude and $\text{c.c.}$ denotes the complex conjugate of the preceding term. Assuming that the wave propagates in the wave direction angle of $\varphi$, the wave phase $\theta$ is  expressed as:

\begin{equation}
\theta =  k_{x} x + k_{y} y - \omega t + \epsilon,
\end{equation}
where, $k_x=k\cos{\varphi}$, $k_y=k\sin{\varphi}$ is the wave number along with the $x$ and $y$ directions, respectively, $\omega$ is the angular frequency and $\epsilon = -(k_{x} x_f + k_{y} y_f)$ is the initial phase at $(\mathbf{X_f},t)=(x_f,y_f, 0)$. $\mathbf{X_f}=(x_f,y_f)$ denotes the position vector of the focused points. Assuming that the wave directions are uniformly distributed over the full angular range $\varphi \in [0, 2\pi)$, the multidirectional wave profile is composed of regular waves with different wave frequencies and wave directions:

\begin{equation}
\eta(\mathbf{x}, t) = \sum_{m=1}^{N_d} \sum_{n=1}^{N_f} \left(a_{mn} e^{i \theta_{mn}} + \text{c.c.} \right),
\label{eq_focused_wave}
\end{equation}
where, the subscript $m$ denotes wave direction component of $\varphi_m$, $n$ wave frequency component of $f_n$. $N_d$ and $N_f$ express the total number of wave direction and frequency components. 
The phase $\theta_{mn}$ becomes zero at the focal point $(x_f, y_f)$ and the focal time $t_f=0$, which ensures that all the constituent waves constructively interfere to form a focused wave $\eta_{*}$.
The wave amplitude $a$ is described using a wave amplitude function $a(f,\varphi)=A(f)D(f,\varphi)$, which is given by a wave amplitude spectrum $A(f)$ and a directional function $D(f,\varphi)$. In this study, we used the Gaussian amplitude spectrum and a uniformly distributed directional function, expressed as follows:

\begin{equation}
A(f) = \frac{\alpha}{\sigma\sqrt{2\pi}} \exp \left( -\frac{\left( \omega-\omega_p \right)^2}{2\sigma^2} \right),
\label{eq_gaussian_amp}
\end{equation}

\begin{equation}
D(f,\varphi)= \frac{1}{2\pi},
\label{eq_dir_func}
\end{equation}
where $\alpha$ is the scaling parameter to adjust the height of the wave at the focused positions $\mathbf{X_f}$ and time of focus $t_f$, which is used in the 'Iterative Amplitude Correction Approach' discussed in Sec. \ref{sec:Amp_correction}. $\omega$ is the wave angular frequency, and $\omega_p$ represents the angular frequency corresponding to the peak frequency $f_p$. 

\begin{figure}[!h]
\centering\includegraphics[width=0.6\textwidth]{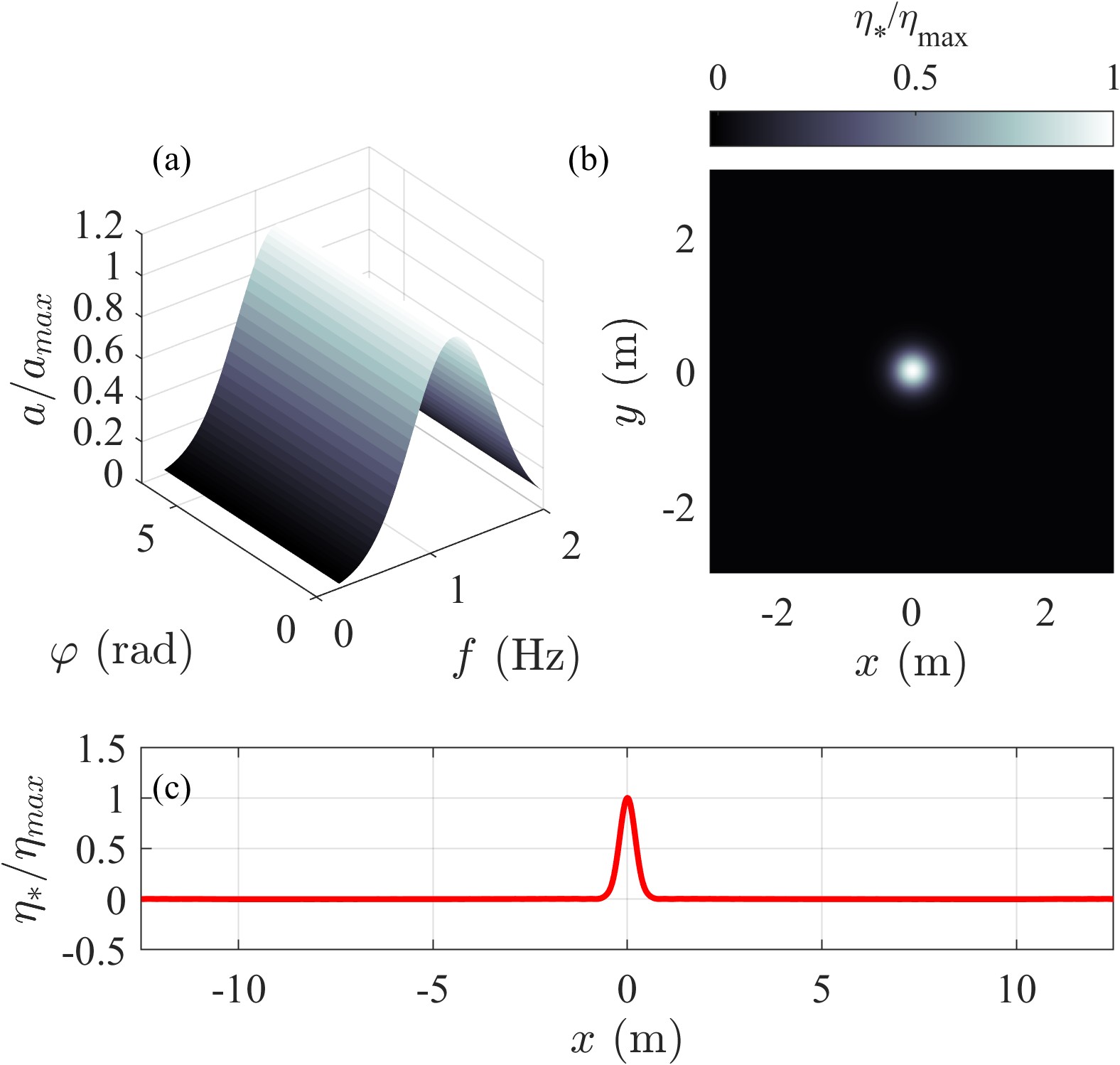}
\caption{An example of focused wave: (a) The Gaussian amplitude directional spectrum, (b and c) Top view of the focused wave and its cross section ($\varphi=0$).}
\label{fig_forcused_wave}
\end{figure}

\begin{figure}[!h]
\centering\includegraphics[width=0.45\textwidth]{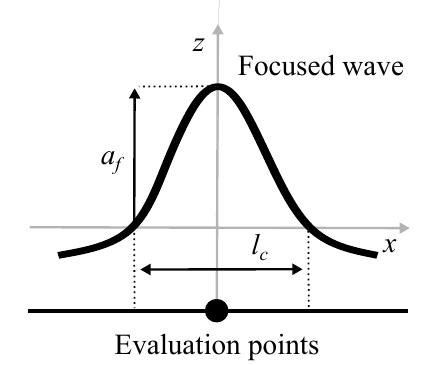}
\caption{Definition of the focused wave, illustrating its amplitude $a_f$ and crest length $l_c$.}
\label{fig_def_forcused_wave}
\end{figure}

In this study, we set $f_p = \omega_p/(2\pi)$ to 1.2~Hz. Increasing $f_p$ results in a more localised focused wave with a reduced crest length $l_c$ defined in \autoref{fig_def_forcused_wave}, thereby minimising interactions with neighbouring evaluation points. This improves the accuracy of the final combined free surface elevation discussed in \S\ref{sec:Amp_correction} and improves the convergence towards the target wave profile ($\eta_{\dagger}$).
The spectral bandwidth $\sigma$ is set to 2.27. As $\sigma$ reduces from this value, the wave profile represented by the red line in \autoref{fig_forcused_wave} exhibits a negative surface elevation on both sides of the focal point ($\mathbf{x_f} = (0, 0)$). In this case, achieving the target surface elevation at a given evaluation point becomes more challenging, leading to poorer convergence.

% Explanation of Figure
% directional spectrum
\Autoref{fig_forcused_wave} illustrates our approach for generating a focused wave at the centre of the basin, where $\mathbf{x_f} = (0, 0)$. Panel (a) presents the Gaussian amplitude directional spectrum.
% final wave profile
By superimposing all frequency wave components while varying $\varphi$ from 0 to $2\pi$, we obtain the final focused wave, represented by red line in panel (c). The spatial distribution of water surface are depicted in panel (b). Notably, the wave profile peaks at the origin and decays toward zero as the distance from the center increases.
This focused wave is generated at each evaluation point, and by adjusting the wave heights at each location to approach the target values, it is possible to generate arbitrary 2D lines and 3D shapes. The following sections introduce the method for creating evaluation points and the final definition of the wave profile.

% ----------------------------------------------------
\subsubsection{Line Integrals Approach using Bezier Curves}
\label{sec:2d_approach}
% ----------------------------------------------------
\begin{figure}[!h]
\centering\includegraphics[width=0.7\textwidth]{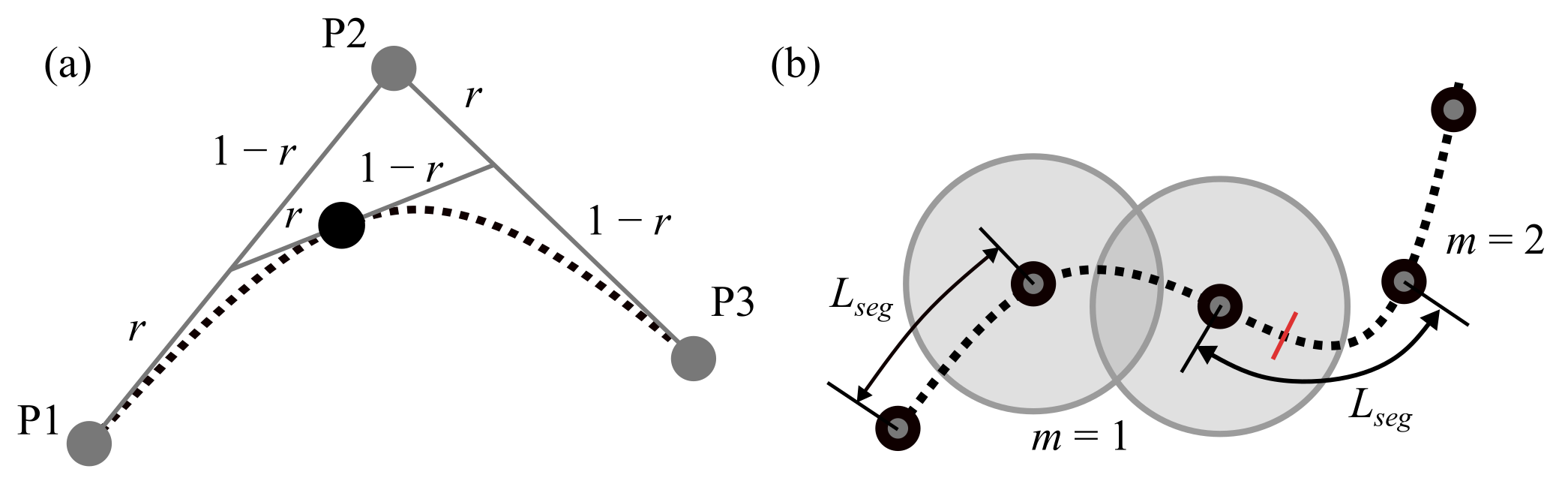}
\caption{Illustration of curve generation and discretisation: (a) Construction of a Bezier curve. (b) Uniform discretisation of two curves, where points are placed at equal arc-length intervals $L_{seg}$ along the curve. The red line represents the boundary between segment $m=1$ and $2$.}
\label{fig_Bezier_curve}
\end{figure}

% Mathmatical explanation of Bizer curves
To represent curves in a mathematically consistent manner, we employ Bézier curves (see, e.g., \cite{farin1993}), which offer a parametric formulation that allows for smooth and flexible shape generation as depicted in \autoref{fig_Bezier_curve}a. A Bézier curve, position vector $\mathbf{C}(r)$ of degree $n$ is defined as:

\begin{equation}
    \mathbf{C}(r) = \sum_{i=0}^{n} B_i^n(r) \mathbf{P}_i, \quad r \in [0,1]
\end{equation}
where the index $i$ ranges from $0$ to $n$, identifying each of the $n+1$ control points,  $\mathbf{P}_i = (x_i, y_i)$ are the control points, and $B_i^n(r)$ are the Bernstein polynomials given by:

\begin{equation}
    B_i^n(r) = \binom{n}{i} (1 - r)^{n-i} r^i.
\end{equation}

\begin{figure}[!h]
\centering\includegraphics[width=0.85\textwidth]{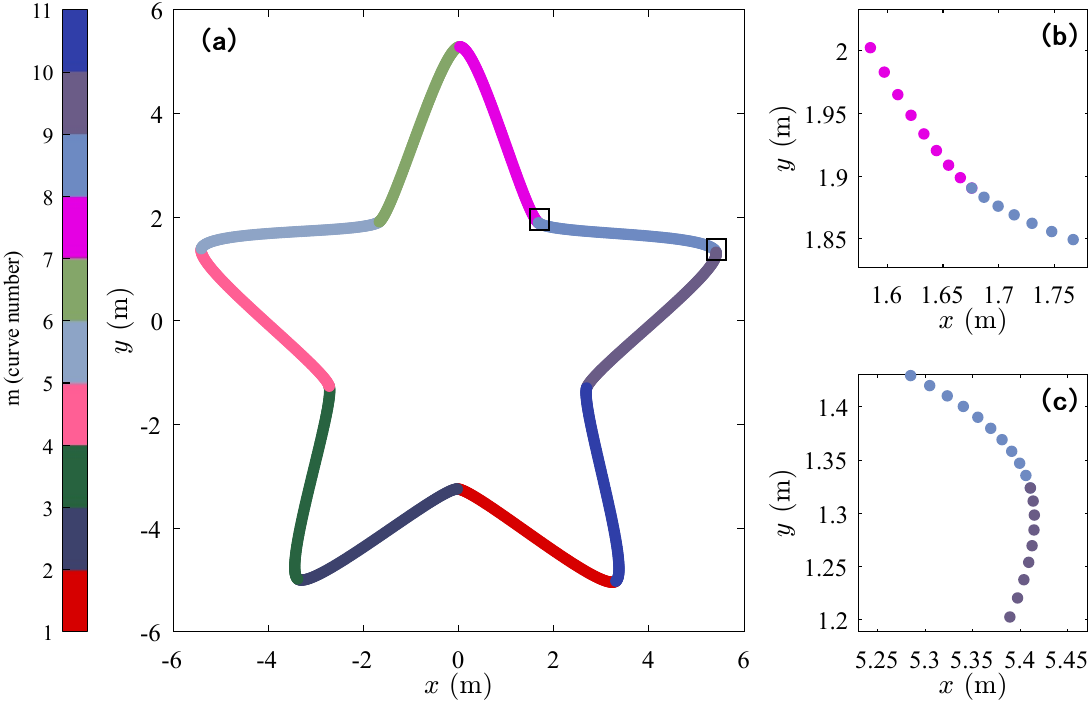}
\caption{Exapmle of Bézier curves and distribution of points on the curves.
(a) The target star shape divided into multiple Bézier curve segments $m$. (b, c) Enlarged views of the boxed regions in (a), demonstrating that evaluation points are more densely concentrated in regions with higher curvature.}
\label{fig_Bezier_example}
\end{figure}

% Discritization of Bizer curves
In our implementation, a target curve is defined as a cubic parametrised curve ($n=3$) using a symbolic variable $r$ in MATLAB, allowing for analytical evaluation of the position, curvature, and arc-length properties. \autoref{fig_Bezier_example}a presents an example plot of the Bézier curves, showing a 2D line drawing of a star consisting of 10 parametric curves. The horizontal ($x$) and vertical coordinates ($y$)  of each curve are discretised over the range $r$ spanning from 0 to 1 with a small increment (e.g. 0.01). The discretisation ensures a smooth representation of the curve by sampling a sufficient number of points along the curve.

% Uniform discretisation of Curves
When applying the aforementioned discretisation method, evaluation points tend to cluster on lines with high curvature as shown in the panels (b and c) in \autoref{fig_Bezier_example}. 
As discussed previously, such clustering of evaluation points can negatively affect the convergence properties of wave amplitude corrections. 
To mitigate this issue, we refine the discretisation process by ensuring that evaluation points are placed at equal arc-length intervals $L_{seg}$ along the parametric curve as depicted in \autoref{fig_Bezier_curve}b. 
The cumulative arc-length $s$ corresponding to the parameter $r_2$ is defined as the cumulative length of the Bézier curve over the parameter interval [$r_1$, $r_2$], given by

\begin{equation}
    s(r_2) = \int_{r_1}^{r_2} \left\| \frac{d\mathbf{C}(r)}{dr} \right\| dr.
\end{equation}

Based on this definition, the corresponding parameter values $r$ are sequentially determined such that the arc length increases by a prescribed segment length $L_{seg}$.
When $s(r_2)$ exceeds $L_{seg}$, the value of $r_2$ is recorded.
By adopting this approach, the distribution of evaluation points along the parametric curve becomes more uniform, thereby improving the accuracy and convergence properties of wave amplitude corrections.

\begin{figure}[!h]
\centering\includegraphics[width=0.96\textwidth]{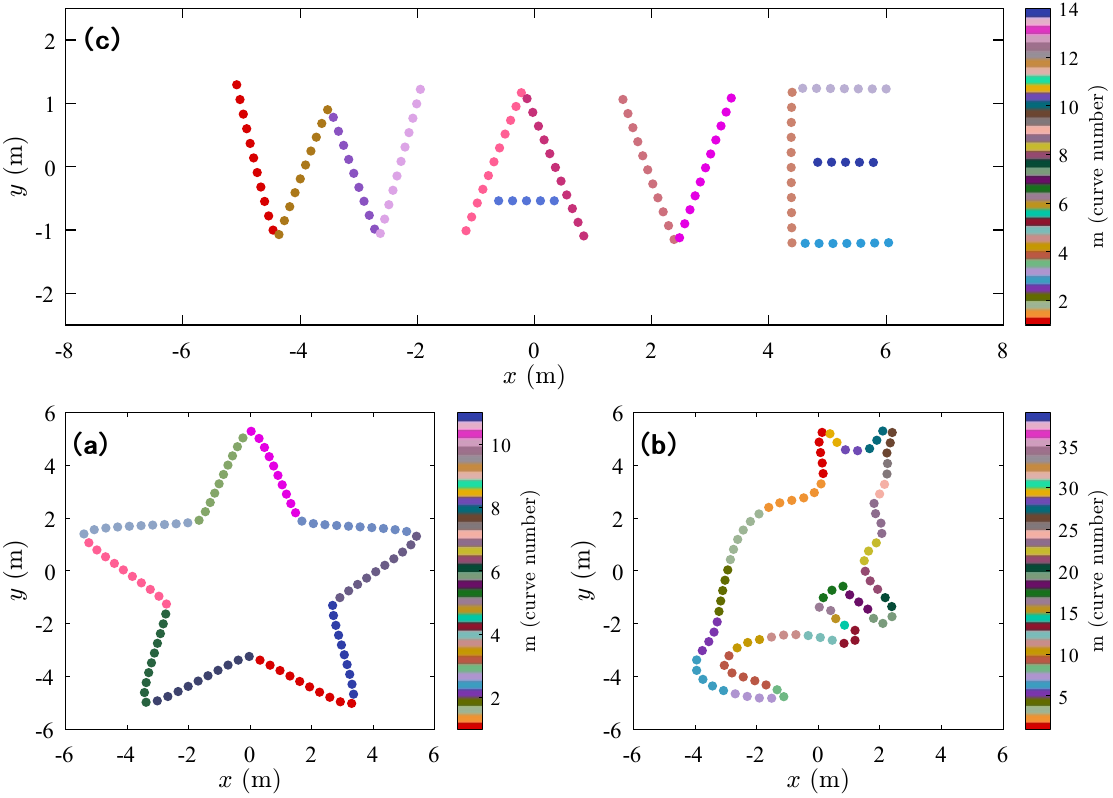}
\caption{An example of evaluation points for two-dimensional space, with a reduced number of segments for improved visual clarity. (a) A star-shaped, (b) a complex profile (cat-shaped), and (c) the word \enquote*{WAVE}. The colours and labels $m$ indicate the curve numbers assigned to different segments of each shape.}
\label{fig_Discrete_Points_2D}
\end{figure}

% Examples of discretised 2D Patterns
\Autoref{fig_Discrete_Points_2D} illustrates three representative 2D patterns used in our study. Panel (a) depicts a star-shaped pattern, while panel (b) presents a cat outline, and panel (c) represents the word \enquote*{WAVE}. Each curve is discretised using the improved uniform arc-length method, ensuring an even distribution of evaluation points. The colours represent different segments of the Bézier curves, demonstrating the segmentation approach used in the uniform discretisation process.
This improved discretisation method enhances the accuracy and stability of wave-amplitude corrections in subsequent analyses as discussed in Sec. \ref{sec:Amp_correction}.

% ----------------------------------------------------
\subsubsection{Discrete Point-Based Approach for Three-Dimensional Shape}
\label{sec:3d_approach}
% ----------------------------------------------------
The approach for generating arbitrary wave shapes can be extended from two-dimensional parametric curves to three-dimensional surfaces. By discretising a given surface into evaluation points, we can control wave amplitudes in a structured manner, allowing for precise wave field generation in three-dimensional space.

To achieve this, evaluation points can be placed using two primary methods:
\begin{itemize}
    \item \textbf{Grid-based placement}: Evaluation points are arranged on a structured grid, which provides a uniform and systematic distribution. This method is particularly useful for defining wave shapes over simple geometric domains such as a Pyramid.
    \item \textbf{CAD-based placement}: Evaluation points are derived from vertex coordinates of a CAD model, allowing for complex and realistic target shapes. This approach enables the generation of intricate wave patterns that conform to complex surface geometries.
\end{itemize}

\begin{figure}[!h]
\centering\includegraphics[width=0.90\textwidth]{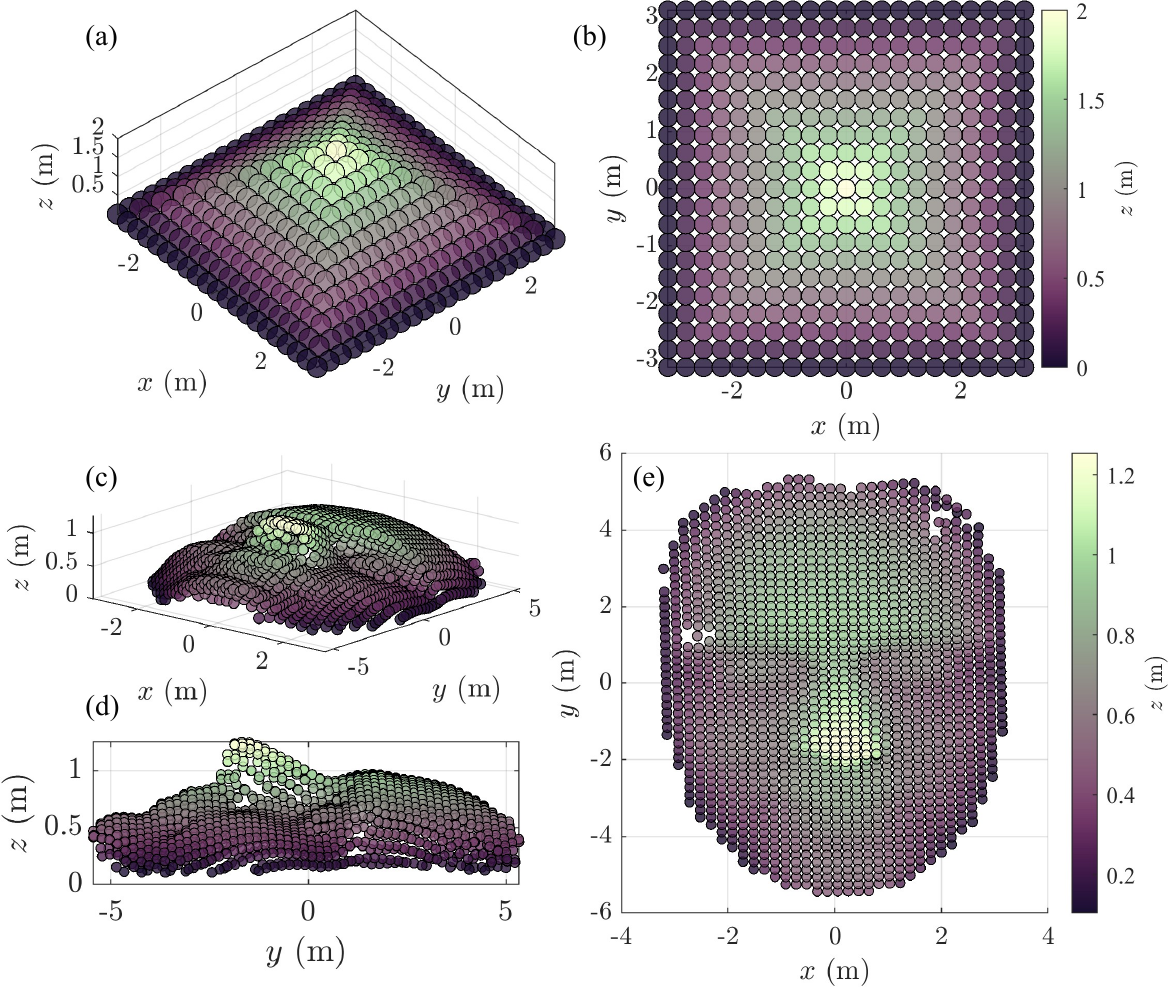}
\caption{An example of evaluation points for three-dimensional space. (a) A 3D perspective view of a Pyramid shape, (b) A top-down view of the Pyramid. (c) A 3D perspective view of points forming a facial shape. (d) A side view and (e) a top-down view of the facial shape.}
\label{fig_discrete_points_3D}
\end{figure}

\Autoref{fig_discrete_points_3D} illustrates an example of evaluation points in three-dimensional space, demonstrating the feasibility of this method for wave shape control. Panel (a,b) shows a Pyramid, and (c--e) presents a face profile obtained from 3D-CAD data in \cite{HESELTINE2008}.

% ----------------------------------------------------
\subsubsection{Amplitude Correction Procedure}
\label{sec:Amp_correction}
% ----------------------------------------------------
As illustrated in \autoref{fig_concept}, the combined wave profile ($\eta$) is defined as the superposition of a focused wave profile $\eta_{l}$ (the subscript $l$ denotes the index of the evaluation point) represented by \autoref{eq_focused_wave}, which can be expressed as:

\begin{equation}
\eta(\mathbf{x}, t) = \sum_{l=1}^{N_p}  \eta_{l}(\mathbf{x},t),
\label{eq_final_wave_profile}
\end{equation}
where, $N_p$ is the total number of evaluation points.
However, if other evaluation points are within the crest length $l_a$, the resultant combined wave may exceed the target wave height. 

To mitigate this issue, a highly efficient 'Iterative Amplitude Correction'  (IAC) approach is introduced into the proposed framework based on linear wave theory. 
At iteration step $n+1$, the correction factor $\alpha^{n+1}$ is formulated as the cumulative product of the ratio of the target water surface elevation $\eta_{\dagger}$ to the combined wave profile $\eta^k$ for each step $k$ from 1 to $n$: 

\begin{equation}
\alpha^{n+1}(\mathbf{x}_f,t_f) = \prod_{k=1}^{n} \frac{\eta_{\dagger}(\mathbf{x}_f,t_f)}{\eta^k(\mathbf{x}_f,t_f)}.
\label{eq_cor_fac}
\end{equation}

For each evaluated point $l$, applying this correction factor $\alpha^{n+1}$ to \autoref{eq_gaussian_amp} yields the updated focused wave profile $\eta_l^{n+1}$ via \autoref{eq_focused_wave}. The updated combined wave field is then obtained by superposition of the focused waves using \autoref{eq_final_wave_profile}.
This approach should iteratively drive the wave profile toward the target.

\begin{figure}[!h]
\centering\includegraphics[width=0.95\textwidth]{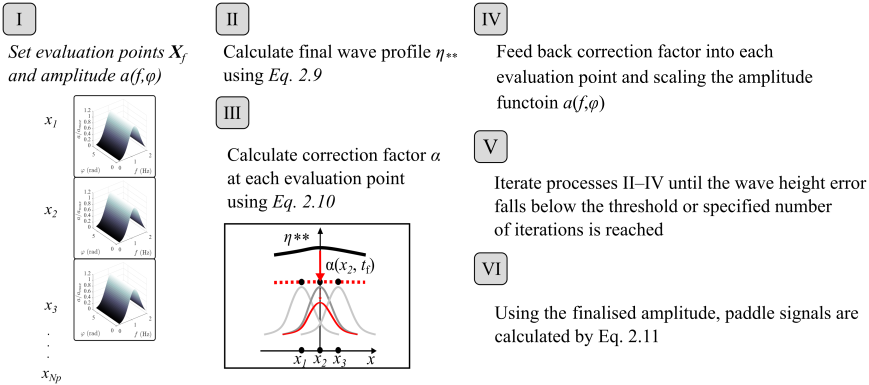}
\caption{Schematic of the method for generating focused wave at the centre of basin and the feedback-based amplitude correction approach.}
\label{fig_amp_correction}
\end{figure}

\Autoref{fig_amp_correction} presents a schematic representation of the methodology for drawing shapes with waves, incorporating a feedback-based amplitude correction approach. The process consists of six sequential steps: ($\mathrm{I}$) defining evaluation points ($\mathbf{x}_f$) and the amplitude function $a(f,\varphi)$; ($\mathrm{II}$) computing the combined wave profile using \autoref{eq_final_wave_profile}; ($\mathrm{III}$) determining the correction factor $\alpha$ at each evaluation point via \autoref{eq_cor_fac}; ($\mathrm{IV}$) applying the correction factor to the amplitude function $a(f,\varphi)$ at each evaluation point; ($\mathrm{V}$) iteratively executing processes II–IV until the wave height error falls below a predefined threshold or a specified number of iterations is reached; and ($\mathrm{VI}$) calculating the paddle signals using the finalised amplitude based on \autoref{eq_wm_signal}. This iterative correction framework ensures that the generated wave field accurately converges to the desired wave profile.

\begin{figure}[!h]
\centering\includegraphics[width=0.95\textwidth]{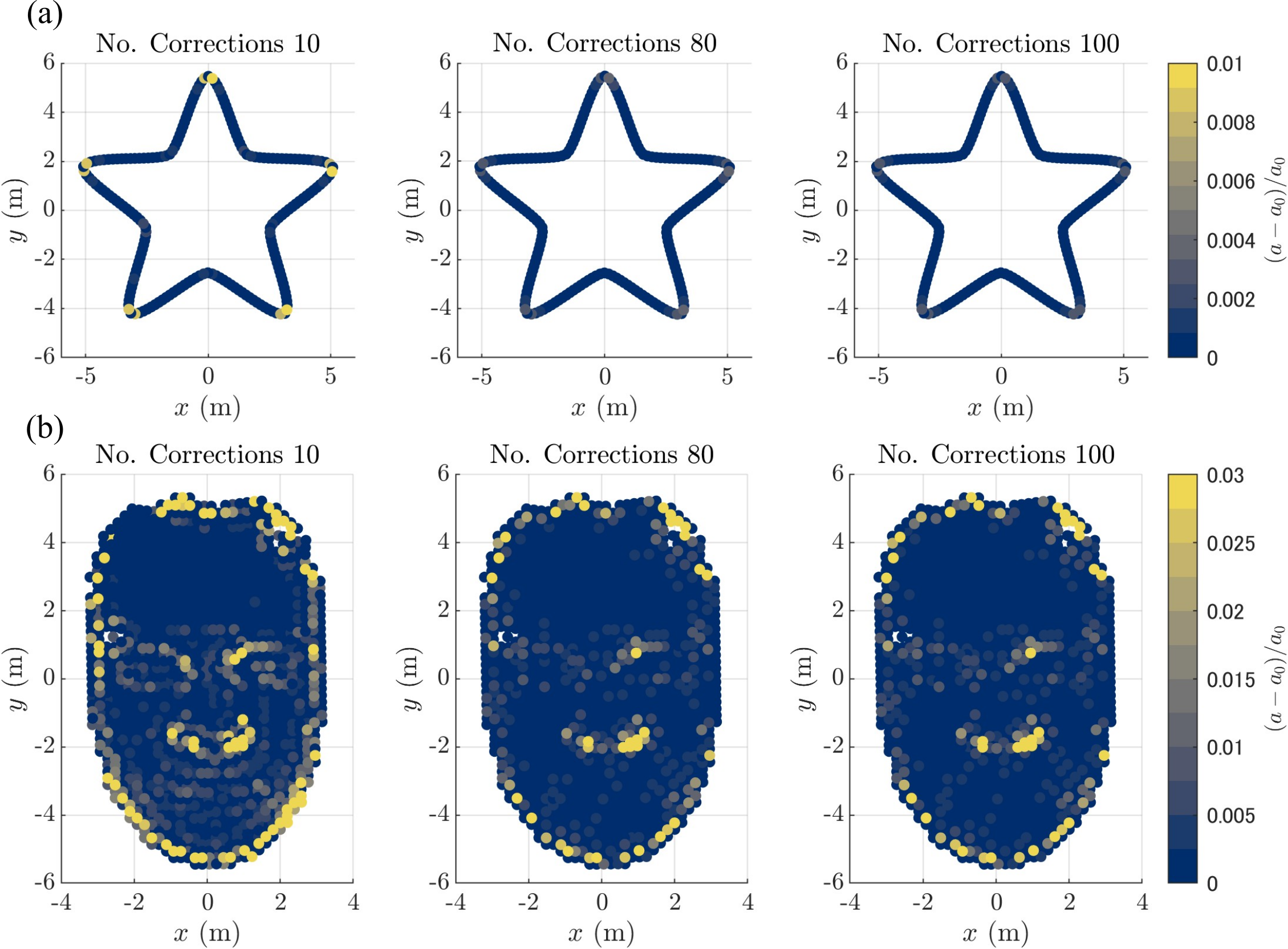}
\caption{Evolution of the relative error in amplitude $(a - a_0)/a_0$ between the corrected amplitude $a$ and the target amplitude $a_0$. The spatial distribution of evaluation points is shown for different numbers of corrections. Panels (a) and (b) show the relative error for the star and face shapes, respectively.}
\label{Amp_Cor_Examples}
\end{figure}

\Autoref{Amp_Cor_Examples} illustrates the spatial distribution of the relative amplitude error, $(a-a_0)/a_0$ , where $a$ is the corrected amplitude and $a_0$ denotes the target amplitude, for different numbers of corrections. Panel (a) displays the relative error distribution for a star-shaped pattern with correction iterations of 10, 80 and 100. Notably, regions with high curvature, such as the tips and inward concave of the star, exhibit slightly elevated amplitude values. This occurs due to the localised overlap of the focused wave. However, the results indicate that as the number of corrections increases the error decreases, leading to a more accurate representation of the desired amplitude.
Panel (b) displays the corresponding results for a 3D face shape, demonstrating a similar trend where the error gradually decreases with additional corrections. The convergence of the relative error was poor at the facial contour due to sparse sampling, and along the nose edges where amplitude gradients were steep. 
However, the results confirm convergence towards the target wave height (less than 3\% error), highlighting the effectiveness of the proposed correction approach.

% ----------------------------------------------------
\subsection{Numerical Simulation and Experiments}
% ----------------------------------------------------
To validate the linear theory for drawing with waves presented in Sec. \ref{sec:focused_wave}, we carry out both numerical simulations and experimental tests. Before conducting experiments we validate our theory numerically using a fully nonlinear particle model configured to be a digital twin of FloWave, as described in \cite{Kanehira19}. Here, we briefly describe the numerical method (\ref{Numerical_1}), boundary condition for wave generation (\ref{Numerical_2}), followed by the experimental setup (\ref{Experimental}). The test cases we conducted are summarised in Table \ref{table_test_cases}, which includes both 2D and 3D cases, as well as simple and complex lines and shapes. 

\begin{table}[!h]
\caption{Test Cases}%%%Table caption goes here
\label{table_test_cases}
\begin{tabular}{llll}%%%The number of columns has to be defined here
\hline
Case & Description & Shape & discretisation Method \\
\hline
1 & 2D simple line & Star & Line Integrals Approach in Sec. \ref{sec:2d_approach} \\
2 & 2D complex line & Cat & Line Integrals Approach in Sec. \ref{sec:2d_approach} \\
2 & Multiple 2D characters & WAVE & Line Integrals Approach in Sec. \ref{sec:2d_approach} \\
3 & 3D simple shape & Pyramid & Discrete Point-Based Approach  in Sec. \ref{sec:3d_approach}\\
4 & 3D complex shape  & Three-dimensional Face & Discrete Point-Based Approach in Sec. \ref{sec:3d_approach} \\\hline
\end{tabular}
\vspace*{-4pt}
\end{table}%%%End of the table

% ----------------------------------------------------
\subsubsection{Numerical model and Set-up}
\label{Numerical_1}
\begin{figure}[!h]
\centering\includegraphics[width=0.95\textwidth]{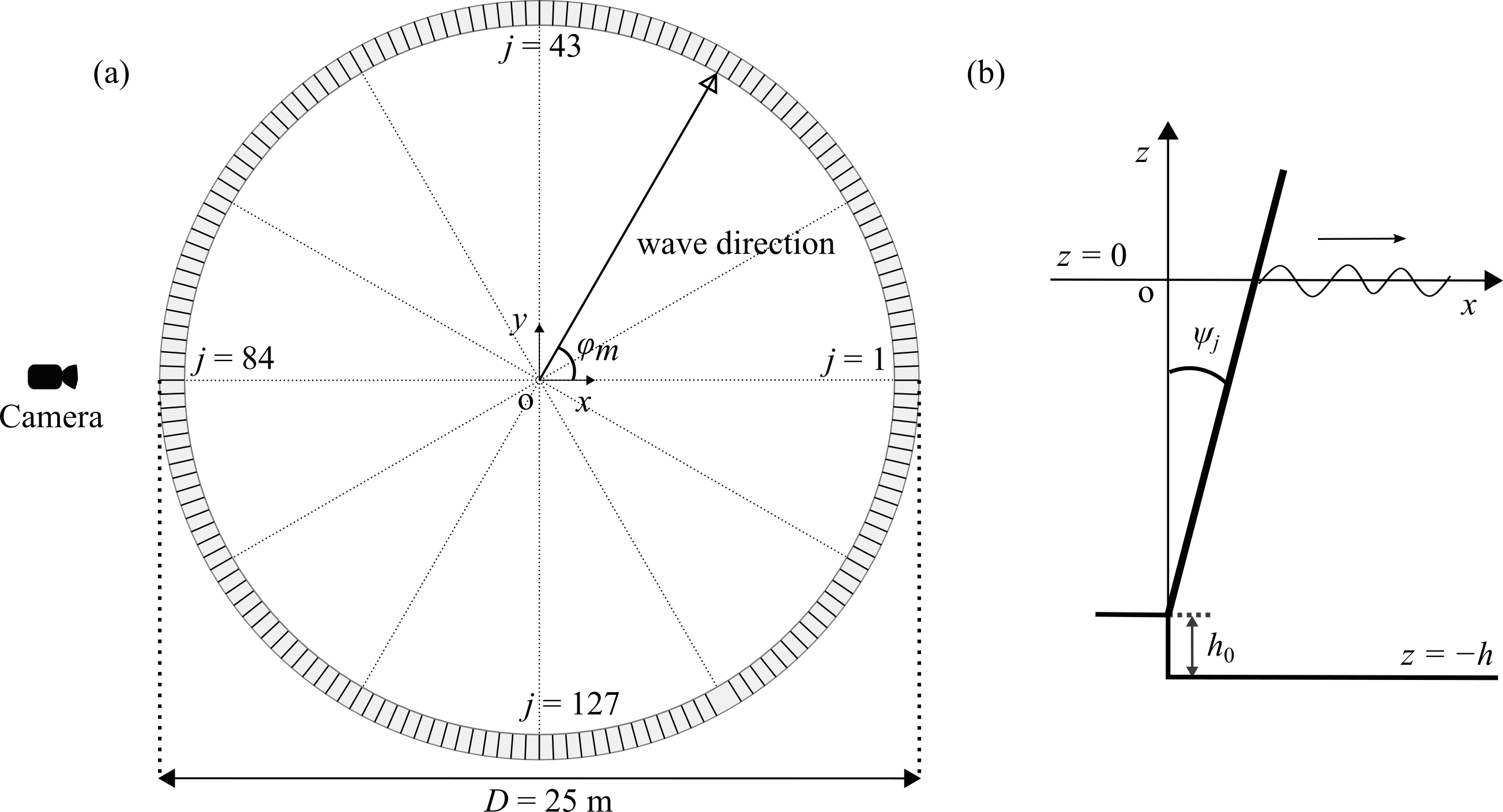}
\caption{Diagram of the FloWave circular wave basin and the hinged flap wave maker. (a) Top-down view of the experimental setup, showing the arrangement of wave makers, wave direction, and measurement devices. (b) Schematic of the wave maker motion in the coordinate system.}
\label{fig_SPH_Basin}
\end{figure}

To simulate wave shapes, and demonstrate realisability in real-world conditions including nonlinearity and dissipation, we utilised a particle-based circular wave basin model \cite{Kanehira19}, as described in \autoref{fig_SPH_Basin}a. The numerical model was developed using the open-source code DualSPHysics and is capable of generating both unidirectional and multidirectional waves. This circular model has no limitation on the angle, enabling the uniform directional distribution to be generated (as described in Sec. \ref{sec:focused_wave}), which is required to make 3D focused waves as shown in \autoref{fig_forcused_wave}b.
Other numerical validations, including the simulation of highly nonlinear directionally spread breaking waves, are presented in \cite{KANEHIRA2020,Kanehira21}.
%  breaf explanation for numerical model
The numerical domain and setup of the basin are illustrated in \autoref{fig_SPH_Basin}. The basin has a diameter of 25 m and a water depth ($h$) of 2 m. The fluid domain is discretised using Smoothed Particle Hydrodynamics (SPH) \cite{Lucy77,Gingold_and_Monaghan_77} with the Wendland kernel \cite{Wendland95}. The fluid motion is governed by the Navier-Stokes equations. 
The pressure of the fluid particles is computed using the Weakly Compressible SPH (WCSPH) formulation, where the Tait equation is employed to relate pressure to fluid density.
% boundary condition
A total of 168 wave makers are positioned around the rim of the circular basin, each consisting of hinged-flap type paddles. Each wave paddle is represented by boundary particles using the Dynamic Boundary Conditions (DBC) \cite{Crespo2007}, and rotates according to an assigned time series of rotation angle data. Further details on the numerical methodology are provided in \cite{Kanehira19}.
All wavemakers and fluid domains are discretised using SPH particles. 
In this study, with a particle initial distance of $d_p$, a resolution of $h/d_p = 100$ 
was employed, resulting in a total of 127.4 million particles.

\subsubsection{Boundary condition for Wave Makers}
\label{Numerical_2}
\begin{figure}[!h]
\centering\includegraphics[width=0.8\textwidth]{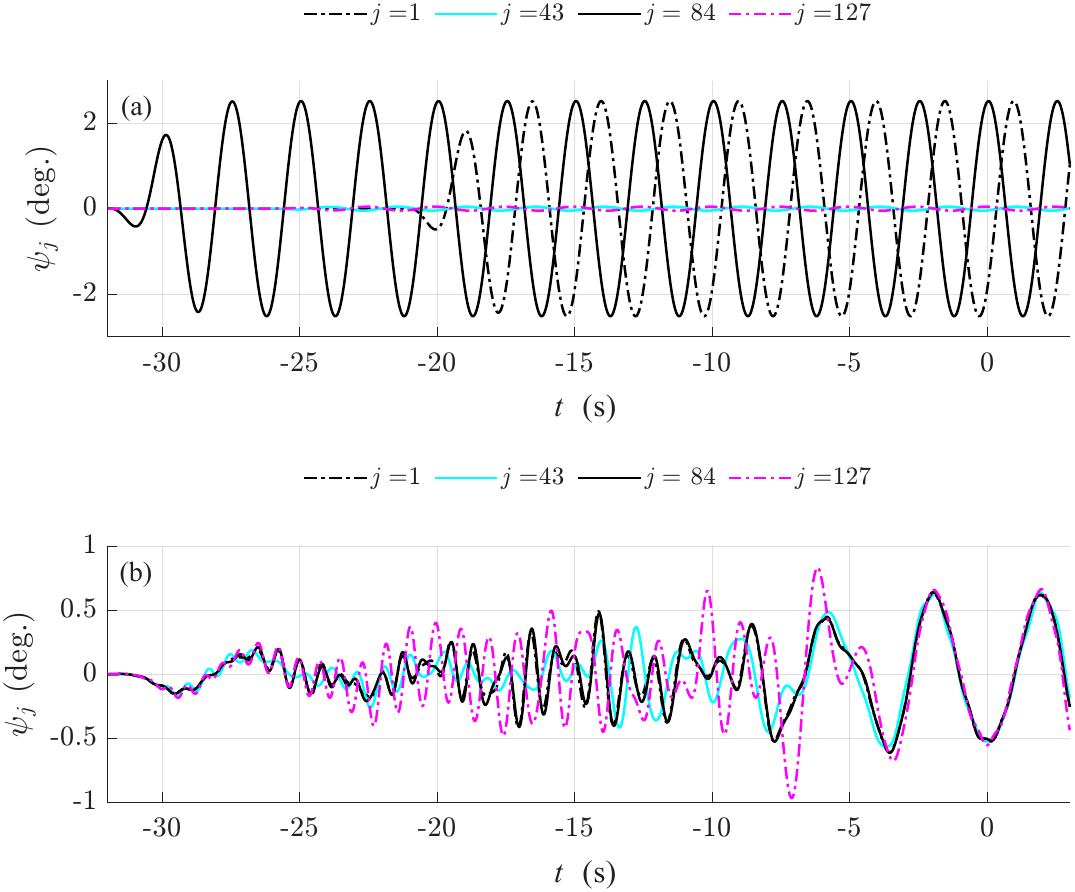}
\caption{Example of the calculated time series of paddle rotation angle using \autoref{eq_wm_signal} for generating (a) long-crested regular wave, (b) the 2D line shape depicted in \autoref{fig_Discrete_Points_2D}a.}
\label{fig_WM_Examples}
\end{figure}

In the previous studies \cite{Kanehira19,KANEHIRA2020,Kanehira21}, the paddle signals obtained from experiments were directly applied to the paddles to accurately reproduce the experimental boundary conditions. In contrast, in this study, to avoid the requirement of physical experiments, we implemented the paddle theory in \cite{GYONGY2014329} for generating regular waves in a circular basin within the SPH model. \Autoref{fig_SPH_Basin}b illustrates the schematic of a flap paddle with an elevated hinge ($h_0$). The paddle angle required to generate long-crested regular waves is given by the following equation:

\begin{equation}
\psi_j (\omega) = - \sum_{l=1}^{N_p}\sum_{m=1}^{N_d} \sum_{n=1}^{N_f} \frac{a_{lmn}(f_n,\varphi_m)}{4k_0 c_0} 
\left[ \frac{\sinh 2k_0 h + 2k_0 h}{\sinh k_0 h} \right] 
\cos \varphi_j e^{-i k_0 R \cos \varphi_j}
\label{eq_wm_signal}
\end{equation}
\begin{equation}
c_0 = \frac{k_0 (h-h_0) \sinh(k_0 h) - \cosh(k_0 h) + \cosh(k_0 h_0)}{k_0^2}
\end{equation}
\begin{equation}
\varphi_{j} = \left( j - \frac{1}{2} \right) \frac{2\pi}{N},\quad j \in \{1,2,\dots,168\}
\end{equation}
% discreption for parameters
where, $\psi$ represents the rotation angle of the paddle, and $\varphi$ denotes the azimuth angle of the paddle. The parameters $k_0$ correspond to the wave number derived from the linear wave theory. $a$ is the amplitude of the regular wave with frequency $f$ and wave direction $\varphi$. The variables $R$ and $h$ indicate the diameter of the basin and the depth of the water, respectively. $N$ denotes total number of paddles of 168, and the subscript $j$ indicates the index of the number of paddles.
% Discreption for equation
The amplitude of the angle of rotation of the paddle is obtained as the real part of the \autoref{eq_wm_signal}, while the initial phase is derived from the imaginary part. \Autoref{fig_WM_Examples} presents examples of the time series of paddle rotation angles, corresponding to wave generation signals for regular waves propagating from left to right in \autoref{fig_SPH_Basin}a.
As the azimuth angle of the paddle $\varphi_m$ increases, the rotation angle of the paddle $\psi_j$ decreases. Specifically, $\psi_j$ becomes zero when $\varphi_m$ equals $\pi/2$ or $3\pi/2$.
\autoref{fig_WM_Examples}b represents the paddle signals for the 2D star shape in \autoref{fig_Discrete_Points_2D}a. 
As shown in \autoref{fig_exp_2D}a, these signals were used as the input signals for the experimental wavemaker.

\subsubsection{Experimental Set-up in the FloWave Circular Wave Basin}
\label{Experimental}

The FloWave Ocean Energy Research Facility, located at the University of Edinburgh in the UK, is a world-leading experimental facility centred around a 25 meter diameter circular test basin \cite{Ingram2014}. The wave generation system consists of 168 active-absorbing force-feedback wavemakers encircling the basin, which enables both the flexible generation of multi-directional waves and the simultaneous suppression of unwanted reflections from the tank boundaries. With a nominal water depth of 2.0 meters in the test area , the circular geometry allows waves to be generated from any arbitrary angle. These advanced control capabilities facilitate high-precision, scaled model testing of complex real-sea conditions, which is essential for evaluating the performance of offshore renewable energy systems (e.g. wave energy converters, tidal stream turbines, offshore wind turbines) and various coastal or offshore engineering structures. In this study, an arbitrary wave field was generated by prescribing time-series torque signals to the 168 wavemakers. These signals are effectively equivalent to the paddle rotation angles calculated using \autoref{eq_wm_signal} once linearly converted to paddle torque.
The operating frequency of these wavemakers ranges from 0.2 to 2~Hz. Unwanted reflected waves are mitigated through an active force-feedback absorption system \cite{Ingram2014}. The focus time was set to 32 s, with a repeat time of 64 s. 
As depicted in \autoref{fig_SPH_Basin}, one camera was set to record the generated wave fields.

% --------------------------------------------
\section{Results}
\label{Sec:Results}
% --------------------------------------------
\subsection{2D lines}

\subsubsection{Wave shapes obtained from linear wave theory}
\begin{figure}[htbp]
\centering\includegraphics[width=0.95\textwidth]{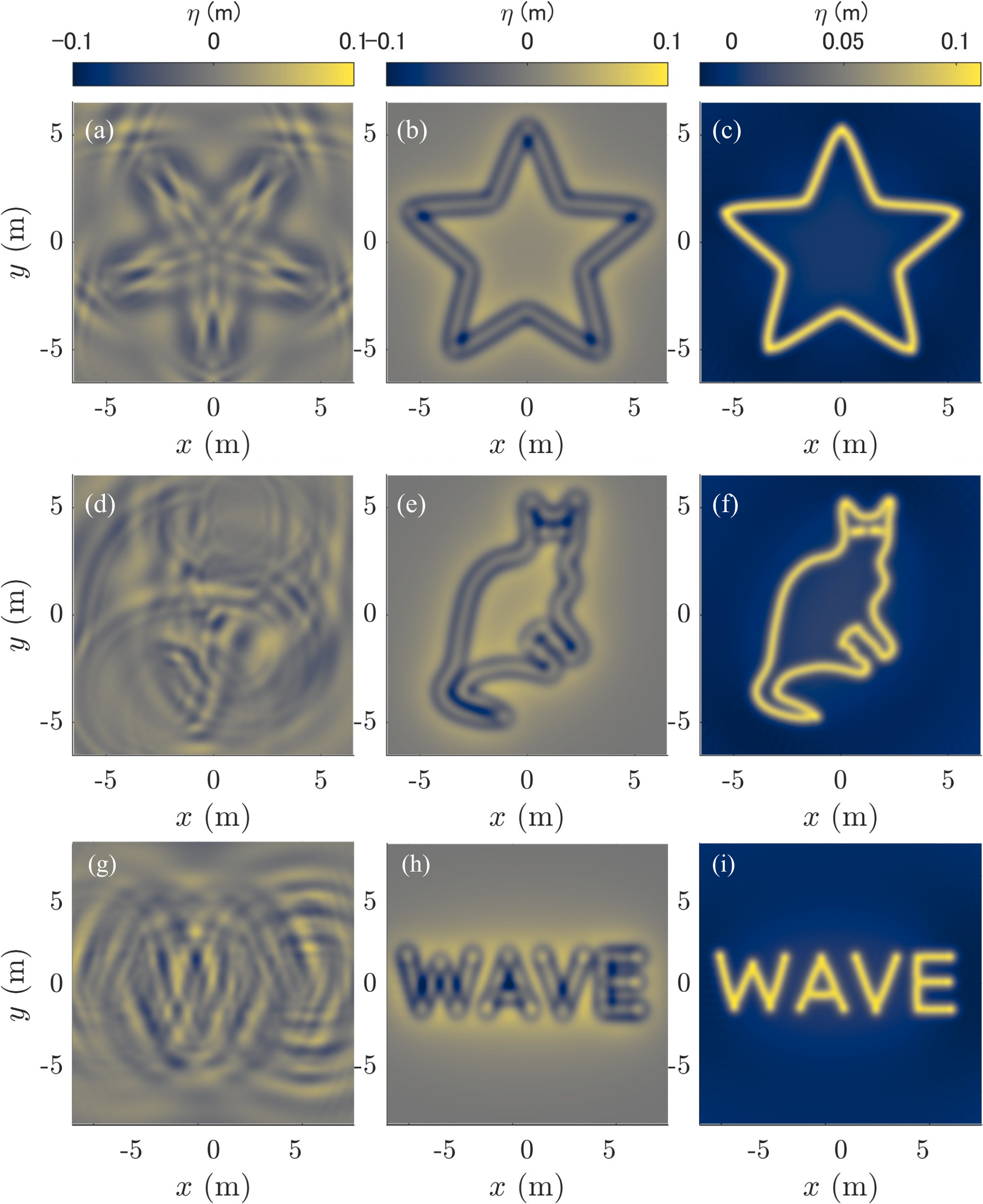}
\caption{Evolution of the simulated wave field at different time steps using linear wave theory, illustrating the formation of the star shape (a--c), 2D cat (d--f) and the Word \enquote*{WAVE} (g--i). From left to right, panels display the wave field over time at $t=-5.4$ s,  $-0.7$ s and $0$ s, respectively.}
\label{fig_lin_2D}
\end{figure}

\Autoref{fig_lin_2D} shows the simulated wave fields using the proposed method based on linear wave theory for the star shape (a--c), the 2D cat (d--f), and the word \enquote*{WAVE} (g--i). From left to right, the panels display the wave fields at three different timestamps, with the rightmost panels (c,f,i) capturing the moment of focusing. Notably, the rightmost panels display sharp boundaries and high contrast in water surface elevation, indicating that the target shapes are reproduced with high spatial resolution. These results validate the robustness of the proposed iterative approach, proving its capability to handle not only simple geometric forms but also intricate patterns and characters by appropriately modulating the amplitude of each wave component.

\begin{figure}[htbp]
\centering\includegraphics[width=0.95\textwidth]{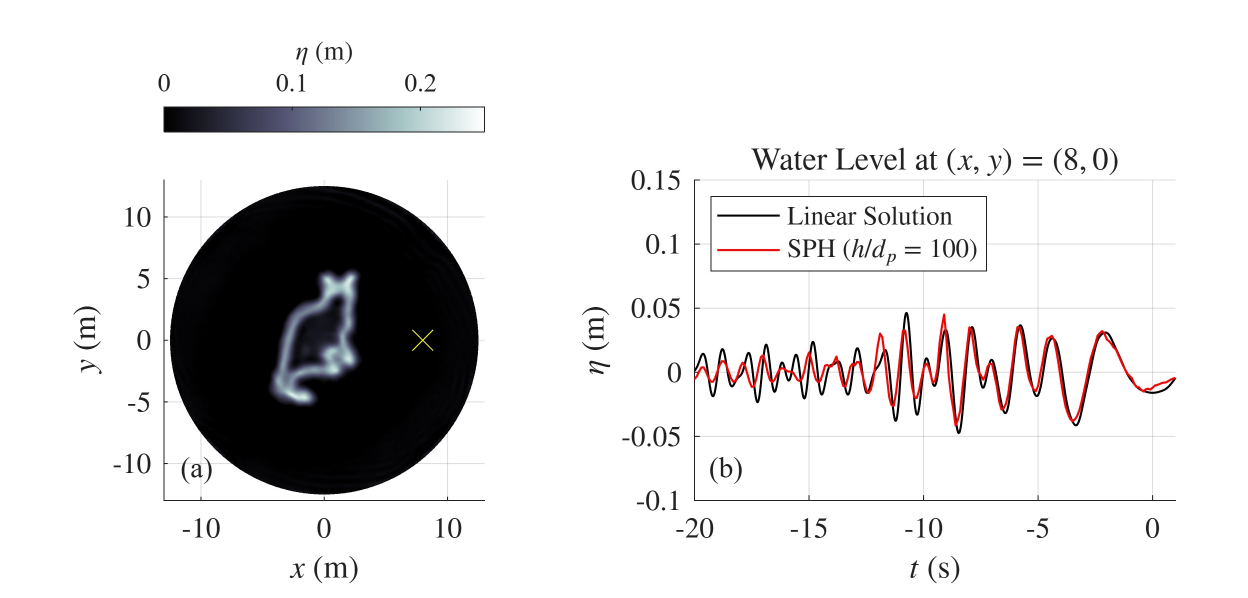}
\caption{(a) Wave field at the time of focus in the SPH simulation, reconstructing the 2D cat. (b) Comparison of the time series of the water surface elevation between linear solution and the SPH result. The yellow cross in panel (a) marks the WG at ($x$,$y$)=(8,0).}
\label{fig_sim_sph_2D}
\end{figure}

\subsubsection{Numerical Validation and Experimental Demonstration}
To validate the effectiveness of the proposed method, numerical simulations using the SPH method were conducted to reproduce arbitrary wave shapes. The boundary conditions for the 168 hinged-flap type wavemakers were derived using \autoref{eq_wm_signal}. Considering the propagation time required for each frequency component to reach the centre of the tank, the focal time was set to $t = 32$ s, consistent with the experimental conditions. \Autoref{fig_sim_sph_2D}a presents the distribution of water surface elevations in the tank at the focal time. As shown, the \enquote*{2D Cat} geometry is clearly formed on the water surface. 
In contrast, \autoref{fig_sim_sph_2D}b compares the time series of the water surface elevation at $(x, y) = (6, 0)$ with the values predicted by linear wave theory. Although large discrepancies are observed between the two, the mismatch during the initial stage ($0 < t < 5$ s) is primarily attributable to the ramp-up effect of the wavemakers, which is not accounted for in linear theory, and undesired waves generated from wavemakers.
To manage computational costs, the particle resolution in this study was set to $H/dp = 100$, which is relatively low compared to previous studies \cite{Kanehira21}. 
Consequently, the pressure and density fields fluctuated during the \enquote{settling process} in which particles initially arranged in a Cartesian grid transition to a stable density distribution through mutual interaction, thereby generating unintended wave ripples. 
Additionally, insufficient resolution leads to the numerical attenuation of high-frequency components as in \cite{Kanehira21}. This is the primary cause of the significant discrepancies in the high-frequency range observed in \autoref{fig_sim_sph_2D}b.
Nevertheless, the recognisable \enquote*{2D Cat} profile validates that \autoref{eq_wm_signal} effectively generates boundary conditions for reproducing complex shapes in circular basins.

\begin{figure}[htbp]
\centering\includegraphics[width=0.85\textwidth]{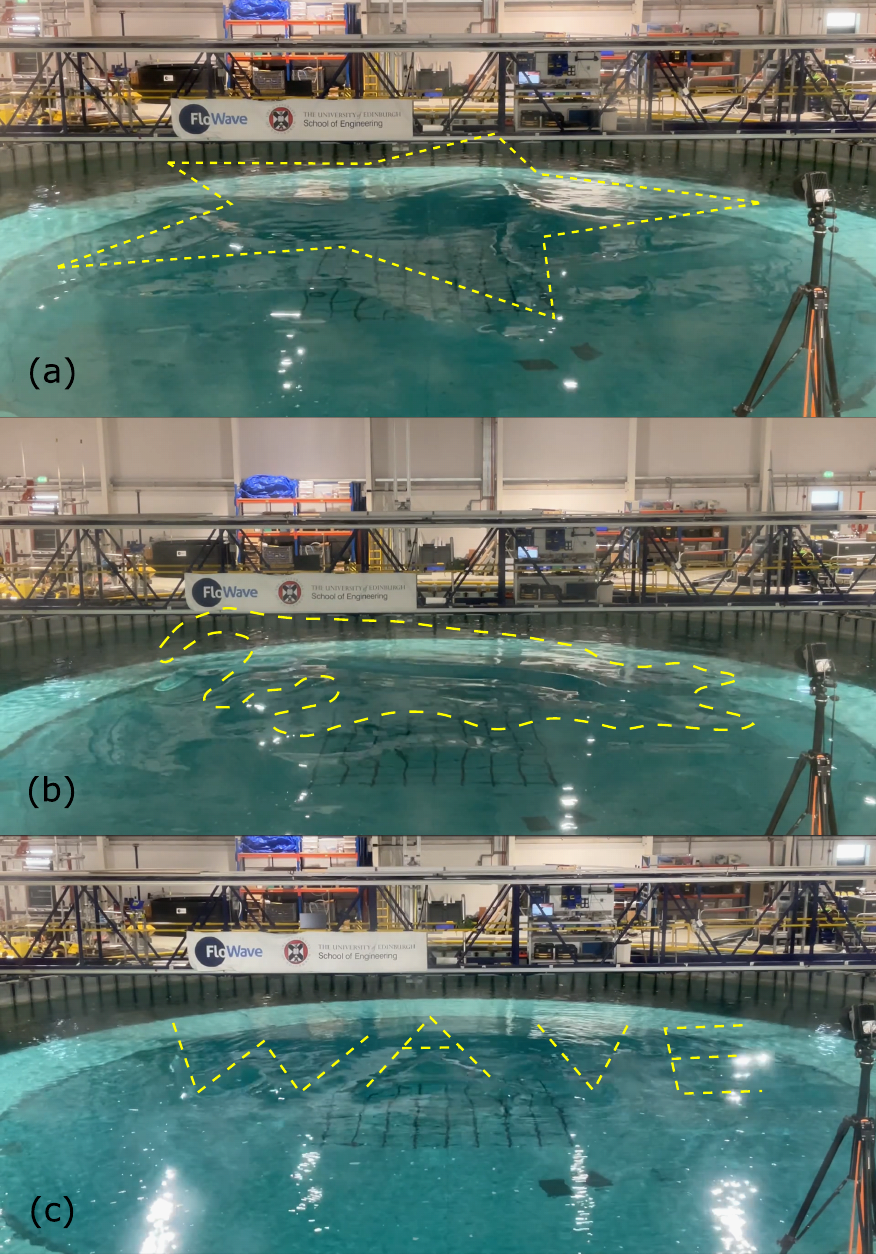}
\caption{Wave fields obtained from experiments in the FloWave tank.
The snapshots show the water surface at the focal time for (a) a star shape, (b) a cat shape, and (c) the word \enquote*{WAVE.} The dashed yellow lines indicate the target profiles for visual comparison.}
\label{fig_exp_2D}
\end{figure}

Following the validation of the boundary conditions via SPH, the same conditions were applied to the wavemakers of FloWave tank. This experiment aimed to verify the physical reproducibility of arbitrary wave fields. \Autoref{fig_exp_2D} presents snapshots of the wave fields near the focal time for (a) the Star, (b) the Cat, and (c) \enquote*{WAVE} shapes. To facilitate visual identification, the target geometries are overlaid as yellow dotted lines.
The Star shape is clearly recognisable in the captured images. However, the more intricate geometries, such as the Cat and \enquote*{WAVE}, proved difficult to identify visually (yet can clearly be generated in practice as evidenced by the numerical model outputs). Several factors likely contributed to this result. First, the lighting conditions were sub-optimal for the photographic measurements. Second, the wave field was disturbed by background noise, including possible residual waves from previous trials and minor reflections from wavemaker boundaries.
%Future improvements should focus on extending the settling time beyond the current 10 minutes. Additionally, increasing the contrast of the 2D contours or thickening the lines by deploying multiple rows of focal points could enhance visibility. Optimising both the geometry selection and the experimental settings remains a subject for future research.

% --------------------------------------------
\subsection{3D shapes}
% --------------------------------------------
\subsubsection{Wave shapes obtained from the Linear wave theory}
\begin{figure}[!h]
\centering\includegraphics[width=0.90\textwidth]{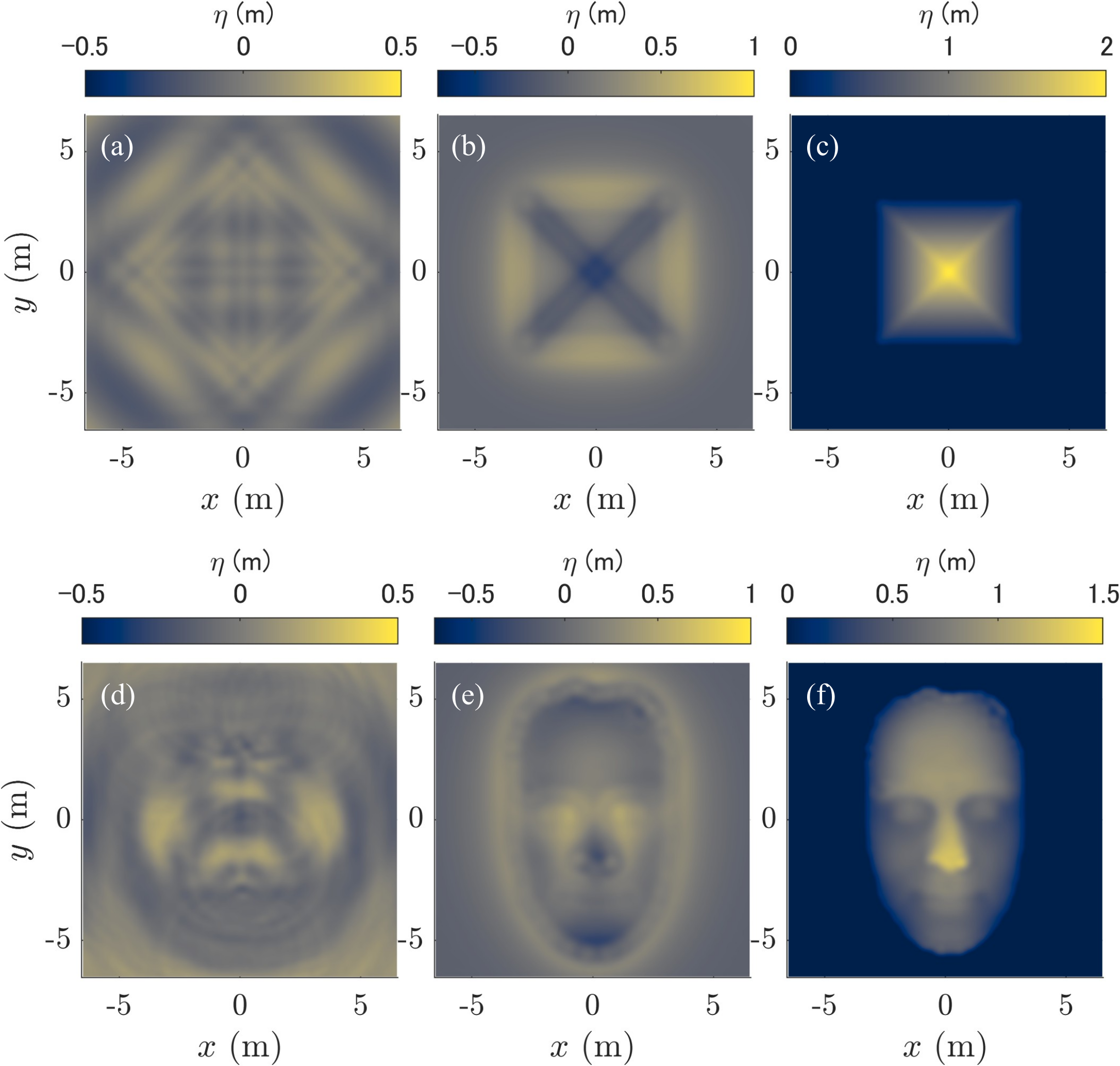}
\caption{Spatio-temporal evolution of simulated three-dimensional wave fields based on linear wave theory. Panels (a–c) and (d–f) illustrate the focusing process of waves into a pyramid geometry and a human face profile, respectively. The snapshots are taken at time steps $t = -5.4$ s, $-0.7$ s, and $0$ s (target time).}
\label{fig_sim_3D}
\end{figure}

\Autoref{fig_sim_3D} presents the numerical simulation results of 3D wave fields using the proposed method based on linear wave theory. Panels (c) and (f) depict the three-dimensional surface elevations of the pyramid and human face profiles, respectively, at the target time ($t = 0$ s). Regarding the pyramid shape in panel (c), localised increases in surface elevation are observed near the edges and the apex. 
Unlike the mutual interference of basis functions at evaluation points previously identified in the 2D cases, these fluctuations are likely attributed to the gradient discontinuities inherent in the target geometry. Analysis of the facial profile in \autoref{Amp_Cor_Examples} further reveals that while the reconstruction achieves high convergence in regions with smooth gradients, such as the cheeks, the accuracy diminishes in areas where the gradient is steep or discontinuous, specifically under the nose and along the facial contours. 
These results indicate that the convergence of the proposed method depends more heavily on the differentiability of the target surface than on the density of evaluation points. Because smooth basis functions cannot perfectly represent features with gradient discontinuities, residual errors tend to concentrate near these edges.
Nevertheless, as shown in panel (f), even the complex facial profile is reproduced with sufficient clarity to be recognisable. This demonstrates the notable versatility and robustness of the proposed method in reconstructing arbitrary and sophisticated three-dimensional wave geometries.

\subsubsection{Numerical Validation and Experimental Demonstration}
\begin{figure}[!h]
\centering\includegraphics[width=0.95\textwidth]{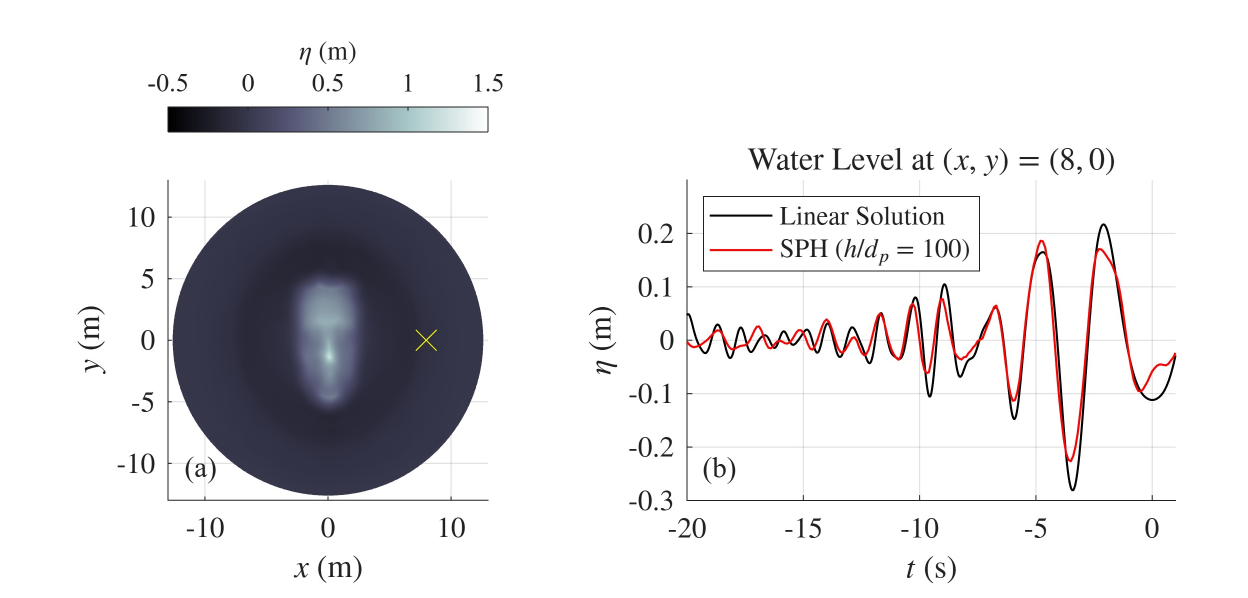}
\caption{(a) Wave field at the time of focus in the SPH simulation, reconstructing the 3D face. (b) Comparison of the time series of the water surface elevation between linear solution and the SPH result. The yellow cross in panel (a) marks the WG at ($x$,$y$)=(8,0).}
\label{fig_SPH_3D}
\end{figure}

The proposed model for reproducing the intended 3D shape was further validated using an SPH simulation. \Autoref{fig_SPH_3D}a illustrates the wave profile of the face generated within the SPH model. For the boundary conditions, predefined time series of rotation angles, calculated from \autoref{eq_wm_signal}, were assigned to all 168 individual wave paddles. \Autoref{fig_SPH_3D}a depicts the wave field of the facial profile at the focusing time of $t=0$ s. It is evident that a discernible facial structure is successfully formed. 
It is worth noting that, compared to the reference profile shown in \autoref{fig_sim_3D}f, the facial details appear globally smoothed and blurred. This effect is attributed to the limited spatial resolution of the SPH model, which causes high-frequency wave components to be dampened and dissipated by numerical viscosity. 
\Autoref{fig_SPH_3D}b illustrates the comparison of water surface elevations between linear wave theory and the SPH result, exhibiting an overall good agreement. Although the wave generation signal in the SPH model was derived from \autoref{eq_wm_signal}, the resulting SPH data at the Wave Gauge (WG) near the wavemaker paddle at $(x, y) = (8, 0)$ align closely with the linear theory time series. Consequently, these results validate the proposed approach for generating three-dimensional waves with arbitrary profiles and verify the accurate implementation of the wave generation theory outlined in \autoref{eq_wm_signal}.
%Indeed, the time-series comparison presented in \autoref{fig_SPH_3D}b confirms the absence of these high-frequency components, although the low-frequency components maintain a relatively high degree of reproducibility.

\begin{figure}[!h]
\centering\includegraphics[width=0.95\textwidth]{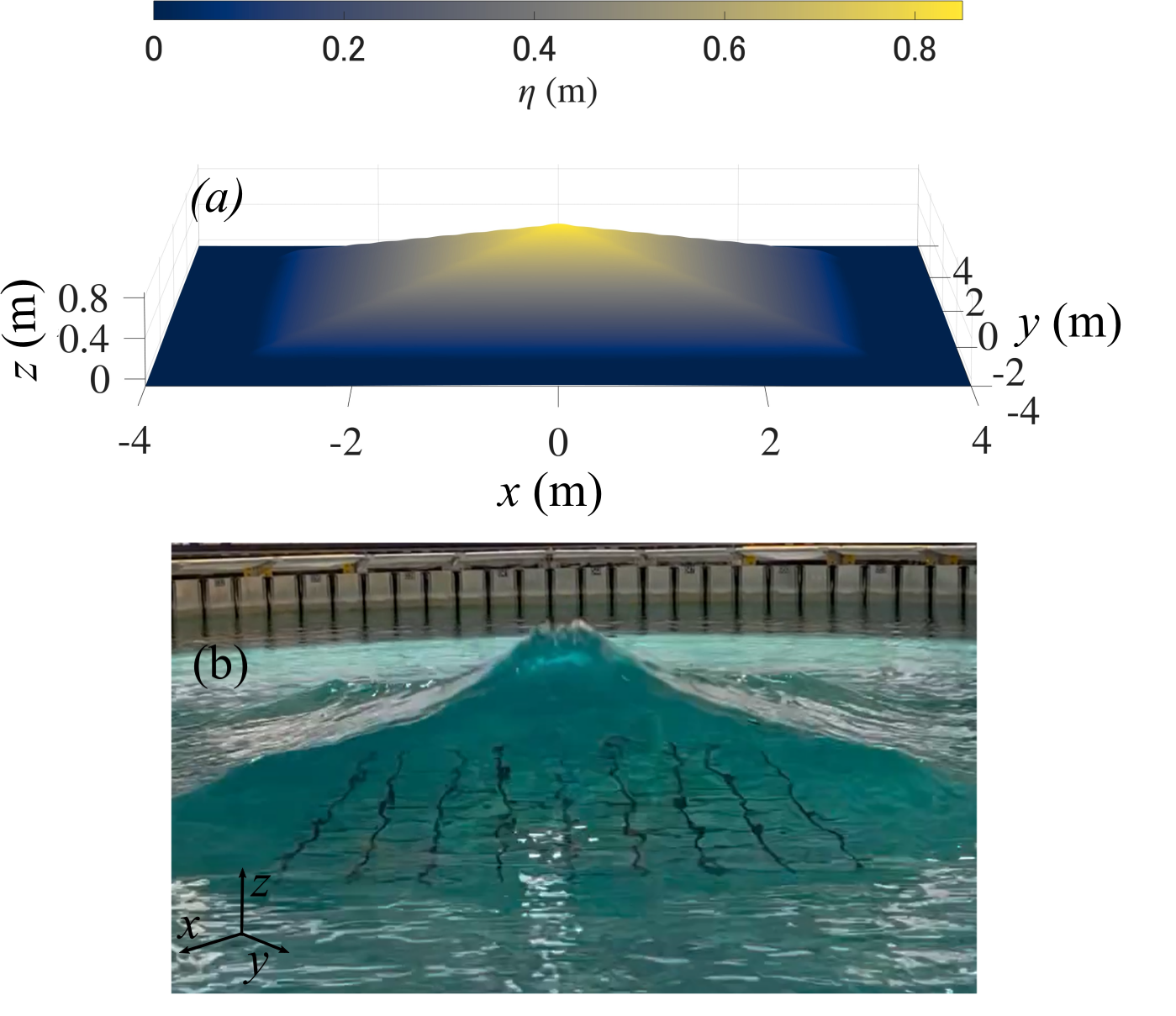}
\caption{Comparison of the 3D pyramid wave profile. (a) Linear wave simulation and (b) experimental observation in FloWave. Nonlinear effects cause the crest to become unstable and approach the breaking point in the experiment.}
\label{fig_exp_3D}
\end{figure}

\Autoref{fig_exp_3D} presents (a) the 3D pyramid shape calculated via linear wave theory and (b) a snapshot of the actual wave generated in the physical experiment, clearly showing its formation at the centre of the wave basin. However, a critical trade-off must be noted regarding the experimental generation of 3D shapes: increasing the target wave height to enhance visual recognisability induces premature wave breaking before the focusing time and introduces significant nonlinear effects into the waveform. Conversely, reducing the wave height to avoid breaking renders the target shape visually indistinct. This phenomenon is evident in \Autoref{fig_exp_3D}b, where nonlinear effects emerge in the central region due to the concentration of high wave energy, resulting in sharpened wave crests similar to those observed in the axisymmetric focused wave in \cite{McAllister22}. Consequently, it was not feasible to experimentally generate the 3D facial shape at a visually recognisable level without triggering wave breaking. Therefore, to accurately predict spatial waveforms while accounting for these nonlinear effects, it is necessary to employ either SPH model simulations as in \cite{Kanehira21} or a higher-order wave theory that incorporates nonlinear interactions for multidirectional waves, such as the third-order formulation by \cite{Madsen_Fuhrman_2012}.

%%%%%%%%%%%%%%% End of first page %%%%%%%%%%%%%%%%%%%%%

\maketitle

% --------------------------------------------
\section{Conclusions}
\label{Sec:Conclusions}
% --------------------------------------------
This study presents a novel methodology for deterministically generating arbitrary 2D curves and 3D volumetric shapes on a water surface using multidirectional focused waves. Distinct from previous approaches based on cylindrical wave expansions \cite{naito2005} or time-reversal approach \cite{ohmatsu2009}, the proposed framework integrates Bézier curve parametrisation with equal arc-length sampling and an Iterative Amplitude Correction (IAC) algorithm. Although our method shares the fundamental concept of representing arbitrary shapes as a set of discrete points---similar to the approach proposed by \cite{ohmatsu2009}---the underlying mechanism for wave generation is distinct. Instead, we utilise a theoretical formulation for generating linear waves within \cite{GYONGY2014329}.
This approach effectively reduces amplitude overshoot caused by wave interference, enabling the precise spatial superposition of spectral components.

The feasibility of the method was first numerically validated through linear wave theory and SPH simulations. These numerical results confirmed the successful reproduction of complex 2D and 3D geometries, verifying the effectiveness of the generated wavemaker signals.

Physical experiments conducted in the FloWave circular basin further demonstrated the generation of target shapes, such as a 2D star and a 3D pyramid. However, the results highlighted a critical trade-off: increasing wave height for better visibility induces premature wave breaking and nonlinear effects. To accurately predict spatial waveforms while accounting for these nonlinear effects, it is necessary to employ either higher-order multidirectional wave theory (e.g. \cite{Madsen_Fuhrman_2012}) or high-resolution SPH simulations \cite{Kanehira21}. Additionally, this technique may hold significant potential for the visual effects and cinematography industries, providing a new deterministic tool for creating realistic and artistic water surface patterns.

%\section{Appendix (Note)}

\enlargethispage{20pt}

\section*{Acknowledgments}
This work was supported by the Japan Society for the Promotion of Science (JSPS) Overseas Research Fellowship.

%%%%%%%%%% Insert bibliography here %%%%%%%%%%%%%%

\vskip2pc

\bibliographystyle{RS}
\bibliography{sample}

\end{document}